\documentclass[prl,twocolumn,amssymb,showpacs]{revtex4}

\usepackage{graphicx}
\usepackage{color}
\usepackage{epsfig}
\usepackage{latexsym}
\usepackage{bm}

\begin{document}

\renewcommand{\ni}{{\noindent}}
\newcommand{\dprime}{{\prime\prime}}
\newcommand{\be}{\begin{equation}}
\newcommand{\ee}{\end{equation}}
\newcommand{\bea}{\begin{eqnarray}}
\newcommand{\eea}{\end{eqnarray}}
\newcommand{\nn}{\nonumber}
\newcommand{\bk}{{\bf k}}
\newcommand{\bQ}{{\bf Q}}
\newcommand{\q}{{\bf q}}
\newcommand{\s}{{\bf s}}
\newcommand{\bN}{{\bf \nabla}}
\newcommand{\bA}{{\bf A}}
\newcommand{\bE}{{\bf E}}
\newcommand{\bj}{{\bf j}}
\newcommand{\bJ}{{\bf J}}
\newcommand{\bs}{{\bf v}_s}
\newcommand{\bn}{{\bf v}_n}
\newcommand{\bv}{{\bf v}}
\newcommand{\la}{\langle}
\newcommand{\ra}{\rangle}
\newcommand{\dg}{\dagger}
\newcommand{\br}{{\bf{r}}}
\newcommand{\brp}{{\bf{r}^\prime}}
\newcommand{\bq}{{\bf{q}}}
\newcommand{\hx}{\hat{\bf x}}
\newcommand{\hy}{\hat{\bf y}}
\newcommand{\bS}{{\bf S}}
\newcommand{\cU}{{\cal U}}
\newcommand{\cD}{{\cal D}}
\newcommand{\bR}{{\bf R}}
\newcommand{\pll}{\parallel}
\newcommand{\sumr}{\sum_{\vr}}
\newcommand{\cP}{{\cal P}}
\newcommand{\cQ}{{\cal Q}}
\newcommand{\cS}{{\cal S}}
\newcommand{\ua}{\uparrow}
\newcommand{\da}{\downarrow}
\newcommand{\red}{\textcolor {red}}

\def\lsim {\protect \raisebox{-0.75ex}[-1.5ex]{$\;\stackrel{<}{\sim}\;$}}
\def\gsim {\protect \raisebox{-0.75ex}[-1.5ex]{$\;\stackrel{>}{\sim}\;$}}
\def\lsimeq {\protect \raisebox{-0.75ex}[-1.5ex]{$\;\stackrel{<}{\simeq}\;$}}
\def\gsimeq {\protect \raisebox{-0.75ex}[-1.5ex]{$\;\stackrel{>}{\simeq}\;$}}


\title{ Thermodynamic Theory of Phase Transitions in Driven Lattice Gases }

\author{ Punyabrata Pradhan and Udo Seifert }

\affiliation{ II. Institut f\"ur Theoretische Physik, Universit\"at Stuttgart,
Stuttgart 70550, Germany }

\begin{abstract}
\noindent{ We formulate an approximate thermodynamic theory of the phase transition in driven lattice gases with attractive nearest-neighbor interactions. We construct the van der Waals equation of state for a driven system where a nonequilibrium chemical potential can be expressed as a function of density and driving field. A Maxwell's construction leads to the phase transition from a homogeneous fluid phase to the coexisting phases of gas and liquid.
  }

\typeout{polish abstract}
\end{abstract}


\maketitle

{\it Introduction.} - Macroscopic properties of systems in equilibrium are described by thermodynamic potentials, like entropy or free energy, which can be derived from the microscopic properties through the Boltzmann distribution. The ultimate triumph of this formalism lies in describing phase transitions, arguably the most interesting phenomena known to occur in various interacting many-particle systems.

A phase transition can also occur in a system with a nonequilibrium steady state (NESS) which exhibits a steady current. However, driven systems have so far resisted attempts to construct a general formalism similar to that in equilibrium \cite{review_thermodynamicsNESS}. Understandably, studies in this direction have focused on getting insights from simple model-systems \cite{Bertini_etal, Derrida_Lebowitz, Hayashi_Sasa2003, Garrido_Goldstein_Lebowitz, Bertin_Dauchot_Droz, Henkes, Wang_Menon}. One such model for systems having a NESS is the driven lattice gas (DLG) \cite{KLS} which has become a paradigm in nonequilibrium statistical physics, analogous to the paradigmatic Ising model or equilibrium lattice gas (ELG). Although the DLG has been studied extensively in the last couple of decades and the various properties concerning the nonequilibrium phase transitions are fairly well known \cite{review_Zia, MFT}, a thermodynamic theory is still missing even for this one of the simplest models of driven interacting many-particle systems.

In this paper, we formulate an approximate thermodynamic theory which not only captures various macroscopic properties but also describes the phase transition in the driven lattice gases with attractive interactions. We construct a mean-field (MF) van der Waals equation of state for a driven system with a chemical potential $\mu(n,E)$ expressed as a function of density $n$ and driving field $E$. The quantity $\mu$ is identified using the concept of  equalization of an intensive variable, like equilibrium chemical potential, for a driven system kept in contact with the corresponding non-driven one. Then we use the Maxwell's construction, familiar in equilibrium for constructing a concave (or convex) thermodynamic potential to describe the phase-coexistence, to explain the phase transition from a homogeneous fluid phase to the coexisting phases of gas and liquid. Our theory is in remarkable agreement with the numerical observations.

{\it Model.} - We consider a model, introduced earlier in \cite{Pradhan_PRL}, of two lattice gases, one driven with volume $V_1$ and the other non-driven with volume $V_2$, exchanging particles through a small contact at $\tilde{V}_1$ and $\tilde{V}_2$, respectively. The energy $H$ of the two systems combined is given by $\nonumber H= K_1 \sum \eta({\bf r_1}) \eta({\bf r_1}') +  K_2 \sum \eta({\bf r_2}) \eta({\bf r_2}')$ where sums are over nearest-neighbor sites with ${\bf r_1}, {\bf r_1}' \in V_1$ and ${\bf r_2}, {\bf r_2}' \in V_2$, $K_1$ and $K_2$ the interaction strengths of the pair-potentials among particles in systems 1 and 2, respectively, and $\eta({\bf r})$ the occupation variable taking values only $1$ or $0$ given the site ${\bf r}$ is occupied or unoccupied, respectively. We choose the jump rate $w(C'|C)$ from a configuration $C$ to $C'$ according to the local detailed balance condition \cite{KLS}: the jump rate for a particle from a site ${\bf r}$ to its unoccupied nearest neighbor ${\bf r}'$ obeys $w(C'|C) = w(C|C') \exp[- \Delta H + E (x'-x)]$ where $\Delta H = H(C') - H(C)$, $E$ is the driving field along the $x$-direction, and $x$ and $x'$ are $x$-components of ${\bf r}$ and ${\bf r}'$ ($k_B T =1$, $k_B$ the Boltzmann constant, $T$ temperature). We consider two-dimensional systems with periodic boundaries in both directions and choose $E=E_1$ when ${\bf r}, {\bf r}' \in V_1$, and $E=0$ otherwise. Also, we confine ourselves to the cases where the combined system is particle-hole symmetric with $K_1=K_2=K$ and consists of particles with attractive interaction of strength $K<0$. For $E_1=0$, the combined system, an equilibrium lattice gas, has the Boltzmann distribution. However, for $E_1 \ne 0$, there is a current in system 1 in the steady state with a probability distribution unknown in general.

{\it Mean-field theory.} - Defining the quantity $w_{\alpha' \alpha}$ as the conditional average of the jump-rate of a particle from system $\alpha$ to system $\alpha'$ if a contact site in system $\alpha$ is occupied and the corresponding contact site in system $\alpha'$ is unoccupied with $\alpha, \alpha'=1,2$ and $\alpha \ne \alpha'$, we get
\be
n_1^{(c)} (1-n_2^{(c)}) w_{21} = n_2^{(c)} (1-n_1^{(c)}) w_{12}
\label{density-rate1}
\ee 
where $n_{\alpha}^{(c)}$ is density at the contact site in system $\alpha$. We define the conditional average, which will be useful later,  
\bea
\langle \hat{\eta}_{\alpha} \rangle_\eta \equiv \sum_{\hat{\eta}_{\alpha}=0}^{4} \hat{\eta}_{\alpha} P(\hat{\eta}_{\alpha}|\eta_{\alpha}^{(c)}=\eta)
\eea
where $\hat{\eta}_{\alpha} \equiv \sum_{i=1}^{4} \eta_{\alpha,i}^{nn}$ is the sum over variable $\eta_{\alpha,i}^{nn}$, the occupation variable at the $i$th nearest-neighbor site to the contact site in system $\alpha$, $\eta=0$ or $1$, and $P(\hat{\eta}_{\alpha}|\eta_{\alpha}^{(c)})$ is the conditional probability of $\hat{\eta}_{\alpha}$ given a fixed value of $\eta_{\alpha}^{(c)}$, the occupation variable at the contact site in system $\alpha$. Now we use a mean-field (MF) approximation for the conditional jump-rate ${w}_{\alpha' \alpha} \approx \exp[-(K_{\alpha'} \langle \hat{\eta}_{\alpha'} \rangle_0 - K_{\alpha} \langle \hat{\eta}_{\alpha} \rangle_1 )/2]$ where the quantity in the round brackets in the exponent is the conditional average of difference in energy between the final and the initial configurations, given that a particle jumps from system $\alpha$ to system $\alpha'$. Therefore we get from Eq. \ref{density-rate1},
\be
\frac{n_2}{1-n_2} e^{{K_2} \left( \langle \hat{\eta}_{_2} \rangle_{_1} + \langle \hat{\eta}_{_2} \rangle_{_0} \right)/2} =  \frac{n_1}{1-n_1} e^{{K_1} \left( \langle \hat{\eta}_{_1} \rangle_{_1} + \langle \hat{\eta}_{_1} \rangle_{_0} \right)/2}. 
\label{MF_eq_state1}
\ee
Here we have implicitly assumed that correlations between the two systems across the contact are negligibly small and consequently $n_{\alpha}^{(c)} \approx n_{\alpha}$ where $n_{\alpha}$ is the bulk-density in system $\alpha$. Finally, putting $K_1=K_2=K$ and then taking logarithm, Eq. \ref{MF_eq_state1} can be rewritten in the more illuminating form, 
\bea
\ln \left( \frac{n_2}{1-n_2} \right) + {\frac{K}{2} \left( \langle \hat{\eta}_{_2} \rangle_{_1} + \langle \hat{\eta}_{_2} \rangle_{_0} \right)}  
\nonumber \\
= \ln \left( \frac{n_1}{1-n_1} \right) + {\frac{K}{2} \left( \langle \hat{\eta}_{_1} \rangle_{_1} + \langle \hat{\eta}_{_1} \rangle_{_0} \right)},
\label{MF_eq_state3}
\eea 
which constitutes the basis of the following analysis. Now one can readily identify the l.h.s of Eq. \ref{MF_eq_state3} as the chemical potential $\mu_2 \equiv \ln [{n_2}/{(1-n_2)}] + {K} \left( \langle \hat{\eta}_{_2} \rangle_{_1} + \langle \hat{\eta}_{_2} \rangle_{_0} \right)/2$ of an equilibrium system, in this MF approximation, with density $n_2$ and interaction strength $K$.

\begin{figure}
\begin{center}
\leavevmode
\includegraphics[width=10.0cm,angle=0]{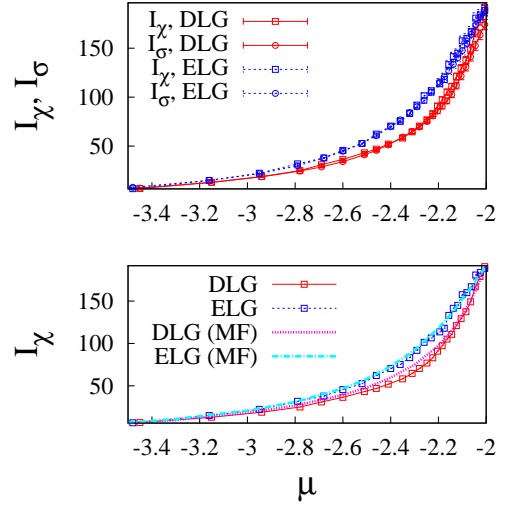}
\caption{Top panel: Simulation results for a $20 \times 20$ system 1 in contact (in the $2 \times 2$ contact region) with a $250 \times 250$ equilibrium system 2, a particle reservoir. Integrated compressibility $I_{\chi} \equiv \int^{\mu}_{\mu_0} ({\partial \langle N_1 \rangle}/{\partial \mu'}) d\mu'$ (squares) and integrated fluctuation $I_{\sigma} \equiv \int^{\mu}_{\mu_0} (\sigma_{N_1}^2) d \mu'$ (circles) as a function of chemical potential $\mu$ for a driven system (red points) with $K=-1$, $E=6$ as well as an equilibrium one (blue points) with $K=-1$ where the systems are in contact with equilibrium reservoirs with the same respective $K$. Bottom panel: The MF results where driven and the corresponding equilibrium case are compared (see main text after Eq. \ref{nnCor_SpecialCase}). All curves obtained in the fluid phase.}
\label{FR1}
\end{center}
\end{figure}

Recent studies of DLGs have revealed a simple thermodynamic structure where, in a large parameter space and to a good approximation, one could define an intensive variable, like equilibrium chemical potential, which equalizes upon contact \cite{Pradhan_PRL}. At this point, we use this concept of assigning the chemical potential $\mu$ of the equilibrium system 2, on the mean-field level, to the driven system 1. We verify it by checking the fluctuation-response relation ${\partial \langle N_1 \rangle}/{\partial \mu} = \sigma_{N_1}^2$ where $\partial \langle N_1 \rangle/\partial \mu$ is the compressibility and $\sigma_{N_1}^2 = (\langle N_1^2 \rangle - \langle N_1 \rangle^2)$ is the fluctuations in particle-number $N_1$ of driven system 1 which is in contact with a particle reservoir equilibrium system 2. This fluctuation relation is indeed remarkably well satisfied as seen in the top panel of Fig. \ref{FR1}.

{\it van der Waals equation of state.} - The above observation leads us to identify the r.h.s of Eq. \ref{MF_eq_state3}, on the mean-field level, as the chemical potential $\mu(n,E)$ for a driven system with density $n$ and driving field $E$, i.e.,
\be
\mu(n,E)=\ln \left( \frac{n}{1-n} \right) + {\frac{K}{2} \left( \langle \hat{\eta} \rangle_1+\langle \hat{\eta} \rangle_0 \right)},
\label{chem_pot1}
\ee  
where we drop the subscript of the occupation variable $\hat{\eta}$. The dependence of $\mu$ on $E$ enters through the conditional average of $\hat{\eta}$ which can be written explicitly in terms of the nearest-neighbor correlation $u \equiv \langle \eta ({\bf r}) \eta ({\bf r}') \rangle$, with ${\bf r}$ and ${\bf r}'$ two nearest-neighbor sites, 
\bea
\langle \hat{\eta} \rangle_{_1} = \frac{\langle \eta^{(c)} \hat{\eta} \rangle}{P(\eta^{(c)}=1)}; \mbox{~~} \langle \hat{\eta} \rangle_{_0} = \frac{\langle (1-\eta^{(c)}) \hat{\eta} \rangle}{P(\eta^{(c)}=0)},
\eea
where $P(\eta^{(c)})$ is the probability of the occupation variable $\eta^{(c)}$ at the contact site. Now using $P(\eta^{(c)}=1) \approx n$, $P(\eta^{(c)}=0) \approx (1-n)$, $\langle \eta^{(c)} \hat{\eta} \rangle \approx 4u_+$ where $u_+ \equiv (u_{\parallel}+u_{\perp})/2$ the average nearest-neighbor correlations in the bulk with $u_{\parallel}$ and $u_{\perp}$ the nearest-neighbor correlations respectively along and perpendicular to the direction of the driving field $E$, and finally defining the average nearest-neighbor correlation function $c_+(n,E) \equiv (u_+ - n^2)$, we obtain from Eq. \ref{chem_pot1},
\be
\mu(n,E) = \ln \left( \frac{n}{1-n} \right) + 2 K M(n, E) 
\label{chem_pot2}
\ee
where the function $M(n, E)$ is defined as 
\be
M(n, E) =\left[ 2n + \frac{c_+(n,E)(1-2n)}{n(1-n)} \right].
\label{def_M}
\ee
The quantities $u_{\parallel}$, $u_{\perp}$ and $u_+$ depend on $E$. Note that, in deriving Eqs. \ref{chem_pot2} and \ref{def_M}, we have used that the density, and the quantities $u_{\parallel}$ and $u_{\perp}$, are approximately equal to those in the bulk. Eq. \ref{chem_pot2} is the desired van der Waals equation of state for a DLG, developed in this paper. When $c_+=0$, i.e., ignoring nearest-neighbor correlations, Eq. \ref{chem_pot2} reduces to the usual MF expression of $\mu$ for an ELG \cite{SKMa}.

\begin{figure}
\begin{center}
\leavevmode
\includegraphics[width=7.2cm,angle=0]{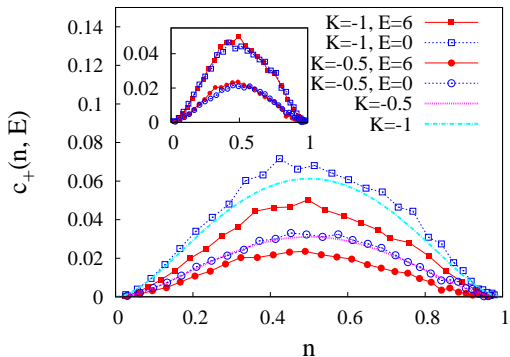}
\caption{Nearest-neighbor correlation function $c_+(n,E)$ for DLG and ELG compared as a function of density $n$ for $K=-1$ and $K=-0.5$. Note that $c_+(n,E) < c_+(n,0)$. The magenta and sky-blue lines are plotted, for $K=-0.5$ and $K=-1$,  using the functional form of Eq. \ref{nnCor_SpecialCase}. Inset: Scaled $c_+(n,0)$ and $c_+(n, E)$ are reasonably well collapsed on each other by using the approximate form of $c_+(n,E) \simeq A(E) c_+(n,0)$ where $A(E=6) \simeq 0.6$ in both the cases.  }
\label{nnCor1}
\end{center}
\end{figure}

\begin{figure}
\begin{center}
\leavevmode
\includegraphics[width=7.8cm,angle=0]{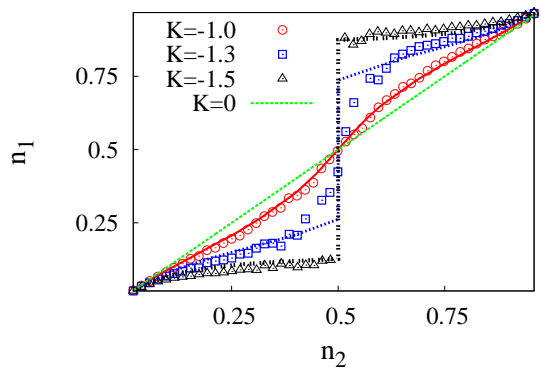}
\caption{Density $n_1$ of a $20 \times 20$ driven system as a function of density $n_2$ of the $250 \times 250$ equilibrium reservoir for various interaction strengths (particle exchange through a $2 \times 2$ contact region). The simulations (points) are compared with the MF theory (lines) which is obtained using a particular choice of $c_+(n,0)$ as given in Eq. \ref{nnCor_SpecialCase} and $A(E)=0.6$, which corresponds to $E=6$, as estimated in Fig. \ref{nnCor1}. Note that,  in the simulations for $K < K_c \simeq -1.30$ as well as in the MF theory for $K < K_c^{\rm{MF}} \simeq -1.22$, there is a jump in the density $n_1$ when the reservoir density reaches $n_2=1/2$.}
\label{n1n2}
\end{center}
\end{figure}

The van der Waals equation of state notably does not have any free parameter. To demonstrate that Eq. \ref{chem_pot2} indeed explains various features of DLGs, we assume a physically motivated approximate form of $c_+(n, E)$ where we use $c_+(n,0) > c_+(n,E)$, for any $n \ne 0,1$, as substantiated in Fig. \ref{nnCor1}. This relation is expected on the ground that the driving field acts as an extra noise to break nearest-neighbor bonds \cite{Bilayr_systems}. For sufficiently small $|K|$, $\mu(n, E)$ is a monotonically increasing function of $n$ (the condition for non-monotonicity is discussed later). For $n<1/2$, since $c_+(n,0) > c_+(n, E)$, the second term in the square bracket of Eq. \ref{def_M} is positive and greater in equilibrium than in nonequilibrium. Since $K<0$, we therefore get $\mu(n,E)>\mu(n,0)$ for a given $n$. It implies that, if a driven system 1 with density $n_1$ is in contact with the corresponding non-driven equilibrium system 2 with density $n_2$ where $n_1,n_2<1/2$, the steady-state densities will be such that $n_1<n_2$. For $n_1,n_2>1/2$, this would be exactly the opposite, i.e., $n_1>n_2$. For $n_1=n_2=1/2$, equalization of the chemical potential $\mu(1/2,E)=\mu(1/2,0)$ implies that densities of a driven and the corresponding non-driven system in contact would indeed be the same. These results, which are derived above using the concept of equalization of a chemical potential for driven systems, are now verified in simulations presented in Fig. \ref{n1n2} where we plot the density $n_1$ of a driven system in contact with the corresponding non-driven system with density $n_2$ for various $K$. Note that they are expectedly consistent with the particle-hole symmetry.

\begin{figure}
\begin{center}
\leavevmode
\includegraphics[width=7cm,angle=0]{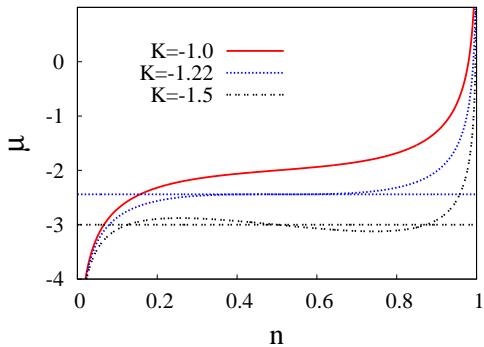}
\caption{The van der Waals equation of state: Chemical potential $\mu$ as a function of density $n$. The curve for $\mu$ develops a kink for $K < K_{c}^{\rm{MF}}(E=6) \simeq -1.22$. The size of the jump in density in the coexistence phase is obtained by using the Maxwell's construction (see the black curve). We have taken $c(n,E)=A(E)c(n,0)$ (see Eq. \ref{nnCor_SpecialCase} and the text below) with $A(E)=0.6$ which corresponds to $E=6$ (see Fig. \ref{nnCor1}). }
\label{Kink1}
\end{center}
\end{figure}

{\it Phase transition.} - Now we describe the phase transition, observed in the simulations, in terms of the thermodynamic potential $\mu$, analogous to the description of the phase transitions in terms of the free energy in equilibrium. As mentioned before, for sufficiently large $|K|$, $\mu$ can be a non-monotonic function of $n$. This is unphysical as one expects that, with increase in the density $n_2$ (or equivalently $\mu_2$) of the equilibrium system 2, the density $n_1$ should also increase. Here, the non-monotonicity of $\mu$ is the signature of the phase transition occurring below a critical value of $K < K_{c}^{\rm{MF}}$. This is verified in the simulations in Fig. \ref{n1n2} where we plot the density $n_1$ of a strongly driven system 1, with $E_1=6$, as a function of density $n_2$ of the non-driven system 2. For $K<K_c\simeq -1.3$, there is a jump in the density $n_1$. The size of the jump goes to zero continuously as $K$ tends to $K_c$ from below. The criticality condition in the MF approximation is given by $(d\mu/dn)= 0$, i.e.,
\be
\frac{1}{n(1-n)} = -2 K \frac{dM}{dn}.
\label{criticality_cond1}
\ee
For any specific form of $c_+(n,E)$, the critical $K_{c}^{\rm{MF}}$ can be found as a solution of $K$ from Eq. \ref{criticality_cond1}. Generically, at $K=K_{c}^{\rm{MF}}$ and $n=1/2$, $(dM/dn)$ has a minimum touching the minimum in the function $1/\{n(1-n)\}$, i.e., the l.h.s. of Eq. \ref{criticality_cond1}. Therefore, writing $[dM/dn]_{n=1/2}=2[1-4c_+(1/2,E)]$, we obtain from Eq. \ref{criticality_cond1} the critical interaction strength $K_{c}^{\rm{MF}}(E) =- 1/[1-4c_+(1/2,E)]$. Now, using $c_+(n,0) > c_+(n,E)$, we get $K_{c}^{\rm{MF}}(E)>K_c^{\rm{MF}}(0)$ which has been observed in our simulations here as well as in simulations in the past \cite{KLS, review_Zia}. Since the equalization of an intensive variable can be related to the existence of a generalized free energy $f(n, E)$ \cite{Bertin_Dauchot_Droz, Pradhan_PRL}, we get $\mu(n,E)=\partial f/\partial n$ and therefore $f(n, E)=\int_{n_0}^{n} \mu(n',E) dn'$. When $\mu$ is a non-monotonic function of $n$, $f(n,E)$ would not be concave. But concavity of $f(n,E)$ could be restored by the usual Maxwell's construction and the jump $\delta n$ in the density can be determined accordingly (see Fig. \ref{Kink1}). Since $c_+(n,E)$ is symmetric around $n=1/2$ due to the particle-hole symmetry, one can see from Eqs. \ref{chem_pot2} and \ref{def_M} that $\mu(n, E)$ is anti-symmetric around $n=1/2$. So the line $\mu'=\mu(n=1/2, E)$ gives equal areas which are bounded by the line and the $\mu(n,E)$ curve. The two density values, where the line $\mu'=\mu(n=1/2, E)$ intersects the curve $\mu(n,E)$ at the left and the right, correspond to the densities in the liquid and gas phase, respectively.

We illustrate the above analysis by taking an approximate form of the equilibrium correlation function 
\be
c_+(n,0) \simeq n(1-n) + \frac{1-\sqrt{1+4 (e^{-K}-1) n (1-n)} }{2(e^{-K}-1)}.
\label{nnCor_SpecialCase}
\ee 
This form can be obtained using approximation of a dynamical mean-field theory \cite{Marro_Dickman}. Then we assume, for simplicity, $c_+(n,E) = A(E) c_+(n,0)$ where $0<A(E) \le 1$ with $A(0)=1$ in equilibrium (see Fig. \ref{nnCor1}). In this special case, using Eq. \ref{chem_pot2}, the chemical potential $\mu$ is plotted as a function of density $n$ for various values of $K$ in Fig. \ref{Kink1} with a specific choice of $A(E)=0.6$ as estimated in Fig. \ref{nnCor1}. Note that the chosen value of $A(E)$ corresponds to a strongly driven system with $E=6$ since the driving field is much larger than the interaction strengths. The kink in $\mu$ appears at $K_{c}^{\rm{MF}}(E=6) \simeq -1.22$ which indicates the onset of the phase transition. We also obtain the equilibrium MF value $K_{c}^{\rm{MF}}(E=0) \simeq -1.62$. Evidently, both values are quite near to the corresponding known values \cite{review_Zia}, $K_c \simeq -1.3$ in strongly driven case and $K_c \simeq -1.76$ in equilibrium. In Fig. \ref{n1n2}, we plot the density $n_1$ of the driven system 1 as a function of the density $n_2$ of the non-driven system 2 where simulations and the MF theory agree quite well, except near the transition region. Moreover, as another validation of the MF theory developed here, in the bottom panel of Fig. \ref{FR1} we plot the integrated compressibility $I_{\chi}$ {\it vs.} chemical potential $\mu$ obtained from the MF theory in the fluid phase with $K=-1$ (i.e., $K < K_{c}^{\rm{MF}}$) where driven and equilibrium cases are compared. The MF results, without any fitting parameter, are again in good agreement with the corresponding simulation results.

{\it Summary.} - In conclusion, we have given an approximate thermodynamic theory which captures various properties of paradigmatic driven lattice gases with attractive nearest-neighbor interactions remarkably well and, in particular, gives a consistent description of the phase transition from the homogeneous fluid phase to the coexisting phases of liquid and gas. Essentially, we have obtained a formalism to calculate a part of a putative nonequilibrium free energy which governs the phase transition observed in various simulations. It still remains to be seen whether and how the long-range correlations observed in these systems \cite{review_Zia, Garrido}, but so far ignored in our analysis, affect this thermodynamic theory of the phase transition.

Importantly, our study opens up the possibility of thermodynamic characterization of driven systems, in general, as the theory can in principle be extended to these systems which exchange a conserved quantity upon contact. However, the challenge in such an extension actually lies in choosing a suitable contact dynamics so that equalization of a thermodynamic variable occurs. 

We thank R. K. P. Zia and R. Ramsperger for discussions.

\end{document}